\documentclass[journal]{IEEEtran}
\IEEEoverridecommandlockouts
\usepackage{cite}
\usepackage{amsmath,amssymb,amsfonts,dsfont}
\usepackage{algorithm}
\usepackage[noend]{algorithmic}
\usepackage{graphicx}
\usepackage{textcomp}
\usepackage{xcolor}
\usepackage{bm}
\usepackage{booktabs}
\usepackage{multirow}
\usepackage{lipsum}
\usepackage{url}
\usepackage{enumitem}
\usepackage{amsthm}
\usepackage{subcaption}
\usepackage{tabularx}
\usepackage[normalem]{ulem}

\AtBeginEnvironment{equation}{\fontsize{10pt}{12pt}\selectfont}
\AtBeginEnvironment{equation*}{\fontsize{10pt}{12pt}\selectfont}

\ifCLASSOPTIONcompsoc
\usepackage[caption=false, font=normalsize, labelfont=sf, textfont=sf]{subfig}
\else
\usepackage[caption=false, font=footnotesize]{subfig}

\newtheorem{theorem}{Theorem}[section]

\newtheorem{remark}{Remark}[section]

\numberwithin{equation}{section}
\renewcommand{\theequation}{\arabic{section}.\arabic{equation}}

\usepackage{geometry}
\title
{
A Survey of Learn-to-Compute Paradigms for \textcolor{black}{Rate-Distortion-Type Problems}
}

\author{
Shitong Wu,
Sicheng Xu,
Lingyi Chen,
Qiang Sun,
Huihui Wu,
Hao Wu,
and Wenyi Zhang, \textit{Senior Member, IEEE}%
\thanks{
Shitong Wu, Sicheng Xu, Lingyi Chen, and Hao Wu are with the Department of Mathematical Sciences, Tsinghua University, Beijing 100084, China.
}%
\thanks{
Qiang Sun and Wenyi Zhang are with the Department of Electronic Engineering and Information Science, University of Science and Technology of China, Hefei, Anhui 230027, China.
}%
\thanks{
Huihui Wu is with the Zhejiang Key Laboratory of Industrial Intelligence and Digital Twin, Eastern Institute of Technology, Ningbo, Zhejiang 315200, P.R. China.
}%
\thanks{Corresponding author: Wenyi Zhang (email: wenyizha@ustc.edu.cn).}
\thanks{This work was partially supported by National Natural Science Foundation of China(Grants 12271289 and 62231022) and by Ningbo Yongjiang Talent Program.}
}

\date{}

\begin{document}
\maketitle

\begin{abstract}
Rate--distortion (RD) theory and its related formulations play a central role in understanding efficient information {representation}, \textcolor{black}{but} computing these quantities remains challenging in high-dimensional settings. Classical iterative methods such as the Blahut--Arimoto algorithm become impractical in {high-dimensional} domains due to the curse of dimensionality and the intractability of mutual-information terms. Recent advances in neural modeling and differentiable optimization offer a promising alternative through a learn-to-compute paradigm, in which probability distributions and objective functionals are represented by flexible neural parameterizations. 
This survey presents an overview of neural approaches for evaluating {the RD-type objectives}. We present three representative families of methods: {variational inference, neural mutual-information estimation, and dual-form optimization.} {By reviewing their theoretical principles, algorithmic techniques, and consistency properties, we elucidate how these methods collectively transform classical RD-type problems into scalable differentiable objectives suitable for deep learning, though challenges remain in large-scale applications.}
 Together, these perspectives offer promising avenues for scaling information-theoretic computation to complex, high-dimensional machine learning systems.
\end{abstract}

\begin{IEEEkeywords}
Rate-distortion theory, Learn-to-Compute paradigm,  Neural estimation, Information bottleneck.
\end{IEEEkeywords}

\section{Introduction}

Machine learning systems are increasingly designed to process high-dimensional sensory data such as images,  videos, and natural language. As the scale and complexity of these modalities continue to grow, the trade-off between preserving essential information and constraining the  resources required for storage, transmission, and computation becomes increasingly pronounced. Rate–distortion (RD) theory ~\cite{shannon1959coding} provides one of the most principled information-theoretic frameworks for characterizing this trade-off, formally specifying the minimum rate required to achieve a given reconstruction fidelity. \textcolor{black}{Several related formulations can be understood within a unified RD framework by adopting different distortion measures. 
For instance, the information bottleneck (IB) principle~\cite{tishby2000information} focuses on extracting task-relevant information, while the indirect rate–distortion (iRD)~\cite{dobrushin1962information} formulation addresses scenarios in which the source signal is observable only through a noisy or incomplete measurement channel.}
These RD-derived formulations have become increasingly valuable in machine learning systems. On the one hand, they provide theoretical foundations for learned image and video compression schemes ~\cite{balle2017end}; on the other hand, they offer conceptual tools for analyzing representation learning and generalization in deep neural networks ~\cite{tishby2015deep}. Collectively, these developments underscore the renewed and growing interest in RD-type problems, which have become central to understanding information processing in high-dimensional regimes of  machine learning.

Several classical works have approached the computation of RD-type functions through iterative optimization in the discrete setting. The Blahut–Arimoto (BA) algorithm~\cite{blahut1972computation,arimoto1972algorithm} is a canonical example: by alternating between updates of the reconstruction rule and the associated marginal distribution, it provides a transparent procedure with guaranteed convergence whenever the underlying alphabets are finite. Subsequent methodological advances have been proposed, including approaches based on the \textcolor{black}{expectation-maximization (EM)} algorithm~\cite{hayashi2023bregman}, constrained-multiplier methods~\cite{chen2025constrained}  and optimal-transport–inspired approaches~\cite{wu2022communication}, each enhancing the convergence speed or enabling \textcolor{black}{RD function} computation under more general constraints in {moderately sized regimes}.
Extending these methods to high-dimensional  settings, however, introduces intrinsic challenges.
{Foremost among these is the curse of dimensionality, where the requisite discretization scales exponentially with the data dimension, rendering classical algorithms computationally  infeasible~\cite{yang2022towards,lei2022neural}.}
Consequently, RD-oriented numerical methods become impractical for the large-scale data modalities encountered in machine learning.

Neural approaches to computational problems have gained significant traction in recent years, offering a scalable alternative to classical numerical methods. 
{Notably, their computational cost grows only linearly with data dimensionality, a far milder scaling than the exponential growth of classical algorithms.}
{Such approaches hinge on two key ingredients: (i) \textit{Parameterized representations of the optimization variables}, and (ii) \textit{Tractable, differentiable estimators of the original objectives}, which together enable end-to-end optimization via stochastic gradient methods~\cite{belghazi2018mutual,lei2022neural}.}
We refer to its application in computational scenarios as the learn-to-compute (LtC) paradigm.
For \textcolor{black}{RD-type problems}, these two ingredients take on concrete forms. On the modeling side, probability distributions can be parameterized by neural architectures, drawing on advances in generative modeling~\cite{kingma2014auto,goodfellow2014generative}. On the objective side, the classical RD formulation—minimizing mutual information under a distortion constraint—must be transformed into a tractable differentiable loss that is compatible with neural parameterization and amenable to gradient-based optimization. This necessitates a reformulation of the RD objective that preserves its information-theoretic structure while enabling feasible {end-to-end neural optimization}.

\textcolor{black}{In recent years, research at the intersection of information theory and machine learning has evolved along two complementary directions: using machine learning to efficiently compute information-theoretic quantities, and using information-theoretic principles to guide the design and analysis of learning models. This survey falls into the former direction. Within this line of work, \cite{czyz2023beyond} investigates neural mutual-information estimation, which is closely related to our goal of making information-theoretic functionals computable in high-dimensional settings. However, it primarily benchmarks neural estimators against classical methods and analyzes their empirical behaviors, whereas our survey focuses on the LtC paradigm and provides a systematic methodological view of neural approaches for RD-type problems. In contrast, the latter direction has also been the subject of several surveys. For example, \cite{goldfeld2020information} reviews how information-theoretic objectives (e.g., IB) are used to guide network design and facilitate the understanding of neural models. Additionally, \cite{yang2023introduction} surveys neural data compression with an emphasis on end-to-end coding architectures, where RD–type objectives are adopted as training losses.}

{This survey provides a comprehensive review of how the objectives arising in RD-type problems can be reformulated into differentiable and computationally tractable forms for neural optimization.} Existing approaches are categorized into three broad formulation families, each offering a distinct mechanism for {addressing} the mutual-information terms and conditional-distribution optimizations that appear across \textcolor{black}{RD-type problems}. The first category comprises variational reformulations, which introduce auxiliary distributions to construct computable relaxations of the original Lagrangians and yield {training objectives analogous to those of variational autoencoders (VAEs)} ~\cite{yang2022towards,alemi2016deep}. The second category is based on neural estimation of mutual information~\cite{belghazi2018mutual}, which replace intractable information terms with trainable critics derived from dual Kullback-Leibler (KL) representations, enabling adversarial or contrastive optimization strategies~\cite{tsur2023rate,zhai2022adversarial}. The third category leverages convex duality to eliminate explicit dependence on the conditional distribution, producing single-variable functionals that can be optimized using neural network parameterizations~\cite{lei2022neural,10806985,11161439}. Together, these three methodological lines provide complementary perspectives on how classical RD formulations can be recast into scalable, differentiable objectives fit for machine  learning systems.

The remaining part of this paper is structured as follows.
Section~II introduces the mathematical notations, reviews the preliminaries of RD theory, and presents a general framework for the LtC paradigm.
Sections~III--V provide detailed expositions of three categories of LtC approaches for solving RD-type problems, covering variational formulations, neural mutual-information estimation methods, and dual-formulation approaches, respectively.
 An empirical
comparison across these methods is summarized and relevant
observations and discussions are presented in Section~VI.
Finally, Section~VII concludes this paper and outlines potential future research directions.

\section{Background}
\label{sec:background}
\subsection{Notations}
\label{sec:notations}

Throughout this paper, we denote random variables by capital letters (e.g., $X, Y, Z$) and their realizations by corresponding lowercase letters (e.g., $x, y, z$). Calligraphic fonts denote the underlying alphabets or spaces (e.g., $\mathcal{X}, \mathcal{Y}$). The probability distribution of a random variable $X$ is denoted by $P_X$, and the conditional distribution of $Y$ given $X$ is written as $P_{Y|X}$. We use $\mathbb{E}_{P}[\cdot]$ to represent the expectation operator taken with respect to distribution $P$. Regarding information-theoretic quantities, $I(X; Y)$ denotes the mutual information between $X$ and $Y$, while $D_{KL}(P\|Q)$ represents the KL divergence between distributions $P$ and $Q$.

In the context of variational learning and approximation, we generally reserve the symbols $Q$ and $R$ (e.g., $Q_{Z|X}, R_{Y|T}$) to denote variational distributions that approximate the true or intractable distribution $P$. Greek letters such as $\theta$, $\phi$, and $\psi$ typically represent the learnable parameters of neural networks. Consequently, a notation such as $Q^\theta_{Z|X}$ indicates that the conditional distribution is parameterized by a neural network with weights $\theta$. Additionally, $\beta$ denotes the Lagrange multiplier governing the trade-off objectives. 

Finally, we consider deterministic mappings parameterized by neural networks. Let $\varphi: \mathcal{Z} \rightarrow \mathcal{T}$ be a measurable mapping and $P_Z$ be a probability measure on $\mathcal{Z}$. We denote the push-forward measure of $P_Z$ by $\varphi$ as $Q_T = \varphi_{\#} P_Z$, which is defined by the property:
\begin{equation*}
    \int_{\mathcal{T}} h(t)Q_T(t)\,dt = \int_{\mathcal{Z}} h(\varphi(z))P_Z(z)\,dz,
\end{equation*}
for any bounded continuous test function $h \in \mathcal{C}(\mathcal{T})$. This notation is particularly relevant when describing generative models where latent distributions are transformed through deterministic neural layers.

\subsection{Rate–Distortion Framework with Generalized Distortions}

The RD theory, established by Shannon~\cite{shannon1959coding}, provides a fundamental characterization of lossy compression under fidelity constraints. 
Given a source random variable $X \sim P_X$, the goal is to determine the minimum information rate (in bits per sample) required to reconstruct $X$ {by a reconstruction variable $Y$} {subject to} an average distortion threshold $D$. 
Formally, the RD function is defined as
\begin{equation}
R(D) = \min_{{P_{Y| X}}:\, \mathbb{E}[d(X,Y)] \le D} I(X;Y),
\label{eq:RD_def}
\end{equation}
where $I(X;Y)$ denotes the mutual information between $X$ and $Y$, and $d(x,y)$ is a prescribed distortion measure such as the mean-squared error. 
This RD function quantifies the theoretical limit of compression efficiency given a desired level of fidelity.

In general, the RD function admits closed-form expressions only for a few special cases~\cite{berger2003rate}. 
For more general distributions, it is useful to consider an equivalent Lagrangian formulation, in which the distortion constraint is incorporated into the objective:
\begin{equation}
\mathcal{L}_{\mathrm{RD}}(\beta)
= \min_{P_{Y|X}} \big[ I(X;Y) + \beta\,\mathbb{E}[d(X, Y)] \big],
\label{rd-lagr}
\end{equation}
where $\beta>0$ is the Lagrange multiplier controlling the trade-off between rate and distortion. 
Geometrically, $\beta$ corresponds to the negative reciprocal of the slope of the RD curve.

\textcolor{black}{Introducing an auxiliary marginal distribution $Q_Y$ and relaxing the mutual information term yields the following equivalent reformulation of \eqref{rd-lagr}~\cite{blahut1972computation,arimoto1972algorithm}:
\begin{equation*}
\min_{Q_Y}
\min_{P_{Y|X}} 
\sum_{x,y} P_X(x)P_{Y|X}(y|x)
\left[\log \frac{P_{Y|X}(y|x)}{Q_Y(y)} + \beta d(x,y) \right].
\label{eq:RD_biconvex}
\end{equation*}
This reformulation preserves the global optimum of the original RD problem. 
It also provides the basis for classical alternating minimization procedures, such as the BA algorithm~\cite{blahut1972computation,arimoto1972algorithm}.}

\textcolor{black}{Beyond its classical formulation, RD theory admits a more general interpretation by allowing task-dependent distortion measures. 
 This flexibility enables the RD framework to capture a broader class of information processing
problems beyond traditional lossy compression.
Two representative examples that fall within this framework are the IB and iRD problems~\cite{tishby2000information, dobrushin1962information}: the IB leverages task-relevant distortion to preserve predictive information, while the iRD addresses scenarios with limited source observability.}

The IB function, introduced by Tishby et al.~\cite{tishby2000information}, extends the classical RD principle from reconstruction fidelity to relevance preservation. 
Rather than optimizing for signal reconstruction accuracy, the IB framework seeks to extract the portion of information in an observable variable $X$ that is the most ``predictive'' of a target variable $Y$.\footnote{{We caution that the symbol $Y$ is used with different meanings in the IB and RD formulations: in IB, $Y$ denotes the target/relevance variable, while in RD it denotes the reconstruction variable. We adopt the IB notation to align with common conventions in most existing literature.}}
This extracted representation is described by a latent bottleneck variable $T$, forming the Markov dependency $Y \!\leftrightarrow\! X \!\leftrightarrow\! T$. 

\textcolor{black}{Under the unified RD framework, the IB problem can be interpreted as an RD problem with a relevance-based distortion~\cite{tishby2000information,goldfeld2020information}. 
Specifically, when the distortion is defined as
\begin{equation}
d(x,t)=D_{\mathrm{KL}}(P_{Y|X=x}\|P_{Y|T=t}),
\label{distortion_ib}
\end{equation}
the resulting objective balances compression and predictive fidelity. 
This distortion corresponds to the logarithmic loss up to an additive constant~\cite{goldfeld2020information}.
Substituting~\eqref{distortion_ib} into the RD objective~\eqref{eq:RD_def} yields the classical IB formulation:
\begin{equation}
      R(I) = \min_{P_{T|X}:\, I(T;Y) \geq I} I(X;T),
\label{ib}
\end{equation}
where $I(X;T)$ quantifies compression and $I(T;Y)$ measures relevance.}

Another notable problem within this unified framework addresses scenarios with limited observability, known as the iRD problem.
Originally formulated by Dobrushin and Tsybakov~\cite{dobrushin1962information}, iRD considers a scenario where the encoder cannot directly access the source variable $S$, but only observes a correlated signal $X$ through a probabilistic channel $P_{X|S}$. 
The goal is to encode $X$ at the minimum rate such that the expected distortion between the original source $S$ and its reconstruction $Y$ does not exceed a threshold $D$. 

\textcolor{black}{Information-theoretic studies~\cite{dobrushin1962information,witsenhausen2003indirect} have shown that the iRD problem can be reformulated as an equivalent RD problem with the induced distortion measure
\begin{equation}
    \bar{d}(x,y)=\mathbb{E}[d(S,y)\mid X=x],
\end{equation}
which captures the expected reconstruction error with respect to the unobservable source. 
Under this formulation, the iRD problem naturally fits into the unified RD framework and takes the standard RD form defined on $X$, leading to
\begin{equation}
R(D) = \min_{P_{Y|X}:\, \mathbb{E}[\bar{d}(X,Y)] \le D} I(X;Y).
\label{distortion_ird}
\end{equation}}


\subsection{The Learn-to-Compute Paradigm}

The LtC paradigm provides a scalable framework for addressing RD-type problems in high-dimensional settings. 
Its central idea is to replace direct optimization over probability distributions with optimization over parameterized model families, while simultaneously replacing intractable information-theoretic objectives by computable differentiable surrogates.

\textcolor{black}{To formalize the LtC paradigm, consider a general optimization problem over probability objects:
\begin{equation*}
\min_{P,\,Q} \; \mathcal{L}(P,Q),
\end{equation*}
where $P$ and $Q$ denote (conditional) distributions, and $\mathcal{L}$ represents a general information-theoretic objective functional (e.g., RD Lagrangian). Such objectives typically involve expectations, divergences, or mutual information terms, which are often intractable in high-dimensional settings.}

\textcolor{black}{Instead of optimizing directly in the space of distributions, LtC introduces parameterized families $\{P_\theta\}_{\theta \in \Theta}$ and $\{Q_\phi\}_{\phi \in \Phi}$, and replaces the original objective with a surrogate:
\begin{equation*}
\min_{\theta,\phi} \; \widehat{\mathcal{L}}(\theta,\phi),
\end{equation*}
where $\widehat{\mathcal{L}}(\theta,\phi)$ is a  differentiable estimator of $\mathcal{L}(P_\theta, Q_\phi)$, typically constructed via stochastic approximation, thereby enabling tractable numerical optimization such as stochastic gradient-based methods. This form of sample-based functional estimation is conceptually an instance of nonparametric estimation~\cite{wasserman2006all} in statistics.
The general procedure of LtC is summarized in Algorithm~\ref{alg:ltc}.}

\begin{algorithm}[ht]
\renewcommand{\algorithmicrequire}{\textbf{Input:}}
\renewcommand{\algorithmicensure}{\textbf{Output:}}
\caption{The Learn-to-Compute (LtC) Framework}
\label{alg:ltc}
\begin{algorithmic}
    \REQUIRE data distribution $P_X$ (accessible via samples only), parameterized model families $\{P_\theta\}$ and $\{Q_\phi\}$, surrogate objective $\widehat{\mathcal{L}}(\theta,\phi)$, learning rate $\eta$, number of iterations $T$
    
    \ENSURE optimized parameters $(\theta^\star, \phi^\star)$ and induced distributions $(P_{\theta^\star}, Q_{\phi^\star})$
    
    \STATE initialize $\theta$, $\phi$
    
    \FOR{$t = 1, \dots, T$}
        \STATE sample $\{x_i\}_{i=1}^n \sim P_X$
        \STATE sample latent or auxiliary variables required by $P_\theta$ and $Q_\phi$
        
        \STATE compute the estimate of the objective:
        \[
        \widehat{\mathcal{L}}_t = \widehat{\mathcal{L}}(\theta, \phi; \{x_i\})
        \]
        
        \STATE update parameters via gradient-based optimization:
        \[
        \theta \leftarrow \theta - \eta \nabla_\theta \widehat{\mathcal{L}}_t, \quad
        \phi \leftarrow \phi - \eta \nabla_\phi \widehat{\mathcal{L}}_t
        \]
    \ENDFOR
    
    \RETURN $(\theta^\star, \phi^\star)$
\end{algorithmic}
\end{algorithm}

\textcolor{black}{
Compared with classical numerical approaches, which typically rely on discretization and iterative updates directly in distribution space, the LtC paradigm shifts the optimization to parameter space and solves it through sample-driven stochastic optimization. 
This shift avoids the exponential discretization burden that arises in high-dimensional settings and makes LtC  suitable for continuous-data regimes encountered in modern machine learning.}

\section{Variational Approaches for Information-Theoretic Estimation}

Variational inference provides a principled approach for approximating complex probability distributions through optimization. 
Among its most influential developments is the VAE ~\cite{kingma2014auto, rezende2014stochastic}. 
The VAE formulates generative modeling as the maximization of a tractable variational lower bound on the data likelihood, known as the \textit{evidence lower bound} (ELBO). 
Specifically, given an observed variable $x$ and a latent variable $z$, the model defines a generative process $P_\theta(x|z)P(z)$ and approximates the intractable posterior $P_\theta(z|x)$ with an encoder $Q_\phi(z|x)$. 
The training objective maximizes
\begin{equation}
    \mathcal{L}_{\text{ELBO}}
    = \mathbb{E}_{Q^\phi(z|x)}[\log P^\theta(x|z)] - D_\mathrm{KL}(Q^\phi(z|x)\,\|\,P(z)),
    \label{VAE}
\end{equation}
which balances reconstruction accuracy with the regularization of the latent space through a KL penalty. 
This formulation not only enables efficient amortized inference via gradient-based learning but also establishes a bridge between probabilistic modeling and neural representation learning.

\textcolor{black}{Subsequent studies have shown that the VAE objective admits an information-theoretic interpretation, with natural connections to RD-type formulations.}
It has been demonstrated that the VAE’s ELBO is mathematically equivalent to the Lagrangian form of the IB objective when the distortion is defined as the negative log-likelihood of reconstruction~\cite{alemi2016deep}. 
Similarly, research on learned image compression has optimized the operational RD objective, revealing its intrinsic connection to the variational autoencoder formulation~\cite{balle2017end}. 
More recently, it has been shown that the RD function itself can be upper-bounded through a parameterized objective resembling the $\beta$-VAE formulation~\cite{higgins2017beta, yang2022towards}, thereby further strengthening the connection between variational latent-variable models and the RD framework.

\textcolor{black}{As a representative example of variational approaches for RD-type objectives, the rate-distortion variational autoencoder (RD-VAE)~\cite{yang2022towards} is trained by minimizing a specific variational objective function.}  This objective, denoted as $J_{RD}$, serves as an upper bound on the classical RD function:
\begin{equation*}
    J_{RD} = \mathbb{E}_{P_X} [D_{\mathrm{KL}}(Q^{\phi}_{Z|X=x} \| Q_Z) ] + \beta \mathbb{E}_{P_X Q^\phi_{Z|X}} [ d(x, \omega(Z))].
    \label{eq:J_RD}
\end{equation*}
In this formulation, $Q^{\phi}_{Z|X}$ represents the encoder parameterized by $\phi$, mapping input $x$ to a latent variable $Z \in \mathcal{Z}$. The function $\omega: \mathcal{Z} \to \mathcal{Y}$ denotes the decoder, which reconstructs the output $Y = \omega(Z)$. The term $Q_Z$ acts as a variational prior, often chosen as a Gaussian, to facilitate the analytical computation of the KL divergence. Notably, this objective is equivalent to a $\beta$-VAE with a likelihood density defined as $P_{X|Z} \propto \exp\{-d(x, \omega(z))\}$.

{From an optimization perspective, the formulation $J_{RD}$ transforms the intractable RD problem into a standard stochastic optimization task. Since the expectations in $J_{\mathrm{RD}}$ are taken over the data distribution $P_X$ and the encoder distribution $Q^\phi_{Z|X}$, they can be efficiently approximated via Monte Carlo sampling. The encoder $\phi$ and decoder $\omega$ can be parameterized by advanced neural architectures such as UNet~\cite{ronneberger2015u} or Normalizing Flows~\cite{rezende2015variational}. Consequently, these parameters can be jointly optimized using stochastic gradient descent (SGD).}

The derivation of this objective stems from a variational reformulation of the classical RD Lagrangian. The standard biconvex RD objective is typically rewritten in expectation form as:
\begin{equation}
\begin{aligned}
    &\mathcal{L}(P_{Y \mid X}, Q_Y) \!=\\&\! \mathbb{E}_{P_X} \![ D_{\mathrm{KL}}\left(P_{Y \mid X=x} \| Q_Y\right)] \!+\! \beta \mathbb{E}_{P_X P_{Y \mid X}}[d(X, Y)].
\end{aligned}
\label{RD-frame}
\end{equation}
By representing $Y$ using a latent variable $Z$ and a decoder $\omega$, and introducing the variational distribution $Q_{Z|X}$, the variational bound $J_{RD}$ is obtained. 

{Next, the connection between the variational formulation $J_{RD}$ and the original RD problem is investigated.}
By treating the latent space $\mathcal{Z}$ as the reconstruction alphabet and defining an induced distortion $\rho_\omega(x, z) = d(x, \omega(z))$, an $\omega$-dependent RD function is defined as:
\begin{equation*}
    R_\omega(D) = \inf_{Q_{Z|X} : \mathbb{E}[\rho_\omega(X,Z)] \leq D} I(X; Z).
\end{equation*}
The relationship between this variational proxy and the true $R(D)$ is formalized by the following theorem:

\begin{theorem}[Theorem~A.3 in \cite{yang2022towards}]
Let $d: \mathcal{X} \times \mathcal{Y} \to \mathbb{R}_{\geq 0}$ be a given distortion measure, and let $\omega: \mathcal{Z} \to \mathcal{Y}$ be a measurable mapping. Define the induced distortion $\rho_\omega(x, z) = d(x, \omega(z))$. Denote by $R(D)$ the classical RD function under $d$, and by $R_\omega(D)$ the RD function under $\rho_\omega$ with $Z$ as the reconstruction alphabet. Then, for all $D \geq 0$,
\[
    R_\omega(D) \geq R(D).
\]
Moreover, if $\omega$ is invertible, then $R_\omega(D) = R(D)$ for all $D$.
\end{theorem}

This theorem guarantees that by minimizing the Lagrangian relaxation of $R_\omega(D)$ (which corresponds to $J_{RD}$), the RD-VAE traces out a curve that upper-bounds the optimal RD trade-off.

\begin{remark}
Recent work such as the Multi-Rate VAE (MR-VAE)~\cite{baemulti}  extends the RD-VAE framework by introducing response hypernetworks. These networks define a function that maps $\beta$ to optimal parameters via hypernetworks, avoiding retraining for each $\beta$. Instead, the hypernetwork learns a continuous mapping $\beta \mapsto (\phi^\star(\beta), \omega^\star(\beta))$, enabling the entire RD curve to be traced in a single optimization run. This approach improves computational efficiency and parameter adaptation smoothness by parameterizing the response of optimal networks to Lagrangian multipliers.
\end{remark}

\textcolor{black}{Following the unified RD perspective established in Sec.~\ref{sec:background}, both the IB and iRD problems can be interpreted as instances of the RD framework with appropriately modified distortion measures. 
Accordingly, their variational implementations can be obtained by adapting the RD-VAE formulation to the corresponding distortions.}

\textcolor{black}{In the IB setting, the distortion measure is replaced by the relevance-based distortion~\eqref{distortion_ib}. 
Compared to standard RD-VAE, this modification changes the role of the decoder: 
instead of reconstructing $X$, the decoder models the conditional distribution $P_{Y|T}$, leading to a predictive  objective.}

Under this distortion, the IB objective admits a variational formulation known as the Variational Information Bottleneck (VIB)~\cite{alemi2016deep}. 
By introducing three variational distributions $Q^\phi_{T|X}$, $Q_T$, and $R^\psi_{Y|T}$, a variational upper bound on the original IB Lagrangian is given by
\begin{equation*}
    J_{\text{VIB}}
    \!=\! \mathbb{E}_{P_X}[D_{\mathrm{KL}}(Q^\phi_{T|X}\|Q_T)]\!
    + \beta\,\mathbb{E}_{P_{X,Y} Q^\phi_{T|X}}[-\log R^\psi_{Y|T}(y|t)].
    \label{eq:J_VIB}
\end{equation*}
Here, $Q^\phi_{T|X}$ is a stochastic encoder mapping $x$ to a compressed representation $T$, 
$R^\psi_{Y|T}$ is a learnable predictor approximating $P_{Y|T}$, 
and $Q_T$ is a variational prior, typically chosen as $\mathcal{N}(0,I)$. 
The second term $\mathbb{E}[-\log R^\psi_{Y|T}(y|t)]$ corresponds to the expected cross-entropy loss, differing from the KL-based distortion in \eqref{distortion_ib} only by an additive constant.

Analogous to RD-VAE~\cite{kingma2014auto}, the objective $J_{\text{VIB}}$ can be optimized via standard VAE-style training using SGD, with expectations estimated through Monte Carlo sampling.

\begin{remark}
Beyond the canonical VIB, several studies have proposed alternative variational approaches for estimating the IB objective. 
Subsequent works~\cite{achille2018information,kolchinsky2019nonlinear} follow the same variational relaxation principle but replace the standard Gaussian prior with richer or task-dependent variational families, enabling more flexible latent representations and potentially tighter approximations.
\end{remark}

\textcolor{black}{For the iRD problem, the distortion is replaced by the induced distortion ~\eqref{distortion_ird}. 
With this choice, the iRD problem can be directly handled within the RD-VAE framework by substituting the standard distortion $d$ with $\bar{d}$.}

\textcolor{black}{When $\bar{d}$ is computable in closed form, the resulting objective can be optimized in the same manner as standard RD-VAE via SGD. 
In data-driven scenarios where $\bar{d}$ must be estimated from samples, additional estimation procedures are required, and specific methodologies have been proposed in~\cite{11161439} (see Sec.~\ref{sec:dual}).}

\section{Neural Optimization Using Mutual Information Estimators}

Beyond variational approaches such as the RD-VAE and VIB, an alternative line of research explores the direct estimation of mutual information without relying on explicit variational relaxations of the original objectives. {Recent advances in machine learning have enabled the development of neural estimators for mutual information.} Among these, the Mutual Information Neural Estimator (MINE) ~\cite{belghazi2018mutual}  stands out as a powerful tool that leverages the Donsker-Varadhan representation of the KL divergence to approximate mutual information in a fully trainable manner. Building upon this foundation, several subsequent works have extended MINE-style mutual information estimation to the computation and optimization of RD-type objectives~\cite{tsur2023rate,zhai2022adversarial}.

\subsection{Mutual Information Neural Estimation}
{The MINE framework~\cite{belghazi2018mutual} provides a neural network–based approach for estimating mutual information between random variables. Specifically, it transforms the estimation of mutual information into a maximization problem over a certain class of functions $T$:
\begin{equation*}
    I(X;Z)
    = \sup_{T}\;
    \mathbb{E}_{P_{XZ}}[T(X,Z)]
    - \log\big(\mathbb{E}_{P_X P_Z}[e^{T(X,Z)}]\big).
    \label{eq:mine-true}
\end{equation*}}

{This representation is derived from the fundamental dual characterization of the KL divergence. Since mutual information is defined as the KL divergence between the joint distribution $P_{XZ}$ and the product of marginals $P_X \otimes P_Z$, i.e., $I(X;Z) = D_{\mathrm{KL}}(P_{XZ}\,\|\,P_X \otimes P_Z)$, it admits the following representation:}

\begin{theorem}[Donsker--Varadhan Representation~\cite{donsker1975asymptotic}]
\label{thm:dv-repr}
For any two distributions $P$ and $Q$ on a measurable space $\Omega$, the KL divergence admits the following representation:
\begin{equation}
    D_{\mathrm{KL}}(P\|Q)
    = \sup_{T:\Omega \to \mathbb{R}}
    \Big\{
        \mathbb{E}_P[T]
        - \log\big(\mathbb{E}_Q[e^{T}]\big)
    \Big\},
    \label{eq:dv}
\end{equation}
where the supremum is taken over all measurable functions $T$ for which both expectations are finite. The supremum is achieved when $T$ satisfies the density ratio condition $dP = e^{T}dQ / \mathbb{E}_Q[e^{T}]$.
\end{theorem}

To obtain a computable and differentiable objective,  MINE parameterizes the function $T$ using a neural network $T_\theta$ with parameters $\theta$. The resulting neural estimator of mutual information is defined as:
\begin{equation}
    \begin{aligned}
        &\widehat{I}_\Theta(X;Z)\\
    =&\sup_{\theta \in \Theta}
    \Big\{
        \mathbb{E}_{P_{XZ}}[T_\theta(X,Z)]
        - \log\big(
            \mathbb{E}_{P_X P_Z}[e^{T_\theta(X,Z)}]
          \big)
    \Big\},
    \end{aligned}
    \label{eq:mine-estimator}
\end{equation}
where the expectations are taken over the joint distribution $P_{XZ}$ and the product of marginals $P_X P_Z$, respectively. This neural estimator is supported by theoretical guarantees, which are formalized in the following consistency theorem.


\begin{theorem}[Strong Consistency of MINE~{\cite[Thm.~2]{belghazi2018mutual}}]
\label{thm:mine-consistency}
Let $\widehat{I}_n(X;Z)$ denote the empirical MINE estimator computed from $n$ i.i.d.\ samples, defined as:
\begin{equation*}
    \widehat{I}_n(X;\!Z)\! = \!\sup_{\theta \in \Theta} \Big\{ \!\mathbb{E}_{P^{(n)}_{XZ}}[T_\theta(X,Z)] -\log\big(\mathbb{E}_{P^{(n)}_X P^{(n)}_Z}[e^{T_\theta(X,Z)}]\big) \!\Big\},
    \label{eq:mine-empirical}
\end{equation*}
where $P^{(n)}_{XZ}$ represents the empirical joint distribution and $P^{(n)}_X P^{(n)}_Z$ is the product of empirical marginals.
Then, for any $\varepsilon > 0$, there exists $N \in \mathbb{N}$ such that
\[
\forall n \ge N,\quad
\big| I(X;Z) - \widehat{I}_n(X;Z) \big| \le \varepsilon, \quad \text{almost surely.}
\]
\end{theorem}

{The proof of this theorem hinges on two key points: (i) the universal approximation property of neural networks, ensuring the estimator can approximate the optimal critic function arbitrarily well; (ii) empirical process theory, guaranteeing almost-sure convergence of empirical expectations to their true expectations.}


The optimization of MINE is performed via stochastic gradient ascent based on empirical expectations. Given sample pairs $\{(x_i, z_i)\}_{i=1}^b$ drawn from the joint distribution and independent samples $\{z'_i\}_{i=1}^b$ drawn from the marginal $P_Z$, the training proceeds iteratively as follows:
\begin{enumerate}[label=(\roman*)]
    \item Compute the empirical estimate of the MINE objective:
    \[
    \widehat{V}(\theta)
    =
    \frac{1}{b}\sum_{i=1}^b T_\theta(x_i,z_i)
    - \log\!\left(
        \frac{1}{b}\sum_{i=1}^b e^{T_\theta(x_i,z'_i)}
    \right).
    \]
    \item Compute the gradient $\nabla_\theta \widehat{V}(\theta)$ via automatic differentiation and update the network parameters:
    \[
    \theta \leftarrow \theta + \eta \,\nabla_\theta \widehat{V}(\theta),
    \]
    where $\eta$ is the learning rate.
\end{enumerate}

\subsection{\textcolor{black}{Applications to Rate-Distortion-Type Objectives}}

The MINE framework can be directly integrated into RD optimization by replacing the intractable mutual information term with its neural estimate, giving rise to a formulation closely related to RD-MINE~\cite{tsur2023rate}. 
To estimate the RD Lagrangian~\eqref{rd-lagr}, the conditional distribution $P_{Y|X}$ is parameterized by a stochastic neural generator $h_\phi$ that maps a source sample $x$ and auxiliary noise $u \sim P_U$ to a reconstruction $y = h_\phi(x,u)$, while the mutual information term $I(X;Y)$ is estimated by a neural critic $T_\theta$ via the MINE objective. 
Substituting these parameterizations yields the min--max optimization problem
\begin{equation}
\begin{aligned}
    \min_{\phi}& \max_{\theta}
    \widehat{\mathcal{L}}_{RD}(\theta, \phi)\\&=\widehat{I}(X,Y;\theta,\phi) \!+\!\beta\mathbb{E}_{P_X P_{Y|X}^\phi}[d(X, Y)],
\end{aligned}
    \label{eq:rd-mine-lagrangian}
\end{equation}
where the mutual information estimator follows~\eqref{eq:mine-estimator}.

\textcolor{black}{Within the unified RD framework, the MINE-based formulation can be extended to other RD-type problems by modifying the distortion measure. For the IB problem, replacing the distortion with the relevance-based distortion measure leads to the objective
\begin{equation}
\begin{aligned}
    \min_{\phi,\psi} &\max_{\theta}\widehat{\mathcal{L}}_{IB}(\theta, \phi, \psi)\\
     =& \widehat{I}(X,T;\theta,\phi) 
     + \beta\mathbb{E}_{P_{X,Y}Q^\phi_{T|X}}\!\big[-\log R^\psi_{Y|T}(Y|T)\big],
\end{aligned}
\end{equation}
which corresponds to the Adversarial Information Bottleneck (AIB) formulation~\cite{zhai2022adversarial}. Here, the parameterization of $R^\psi_{Y|T}$ follows that in the VIB formulation. Similarly, the iRD problem can be handled by replacing the distortion $d$ with the induced distortion $\bar{d}$.}

Such min--max problems derived from the MINE framework are typically solved via nested alternating optimization~\cite{tsur2023rate, zhai2022adversarial}. Taking the RD objective~\eqref{eq:rd-mine-lagrangian} as an example, in each outer iteration, multiple critic updates are first performed with the generator parameter $\phi$ fixed to obtain a more accurate mutual-information estimate. Subsequently, with the critic fixed, the generator is updated to improve the encoder (i.e., the conditional distribution). This nested procedure induces an adversarial dynamic analogous to Generative Adversarial Networks (GANs)~\cite{goodfellow2014generative}. The overall procedure is summarized as follows:

\begin{enumerate}[label=(\roman*)]
    \item Sample mini-batches $\{x_i\}_{i=1}^m \sim P_X$ and noise $\{u_i\}_{i=1}^m \sim P_U$, and compute the joint reconstructions $y_i = h_\phi(x_i, u_i)$.
    
    \item Generate $\{y'_i\}_{i=1}^m$ by shuffling $\{y_i\}$ to approximate the marginal distribution $P^\phi_Y$ independent of $P_X$.
    
    \item Compute the empirical MINE estimator:
    \[
        \widehat{I}_m(X,Y;\theta,\phi)
        = \frac{1}{m}\sum_{i=1}^m T_\theta(x_i, y_i)
        - \log \left( \frac{1}{m} \sum_{i=1}^m e^{T_\theta(x_i, y'_i)} \right).
    \]
    
    \item Inner critic-update loop: with $\phi$ fixed, update $\theta$ via gradient ascent for $K$ steps to maximize the mutual-information estimate:
    \[
        \theta \leftarrow \theta + \eta_\theta \nabla_\theta \widehat{I}_m(X,Y;\theta,\phi).
    \]
    
    \item Outer generator-update step: with $\theta$ fixed, update $\phi$ via gradient descent to minimize the total Lagrangian:
    \[
        \phi \leftarrow \phi - \eta_\phi \nabla_\phi \Big( \widehat{I}_m(X,Y;\theta,\phi) + \beta \frac{1}{m}\sum_{i=1}^m d(x_i, y_i) \Big).
    \]
\end{enumerate}

\begin{remark}
RD-MINE~\cite{tsur2023rate} directly estimates the constrained RD function~\eqref{eq:RD_def} by incorporating a differentiable projection mechanism that enforces the constraint $\mathbb{E}[d(X, Y)] \le D$. 
Specifically, an intermediate latent variable $Z_i = g_\phi(X_i, U_i)$ is generated and projected to the final reconstruction via
\[
    Y_i = X_i + \frac{D}{\hat{d}(X^n, Z^n)}(Z_i - X_i),
\]
where $\hat{d}(X^n, Z^n)$ denotes the batch-wise average distortion. 
This operation maps the generator output onto the distortion constraint boundary, making the constrained problem amenable to neural optimization.
\end{remark}

\section{Dual-Formulation Based Neural Methods}
\label{sec:dual}

Besides variational and mutual information estimation approaches, 
another prominent line of work explores dual representations of the RD-type problems~\cite{lei2022neural,10806985,11161439}.  
Recent studies have sought alternative formulations that avoid explicit variational bounds 
and instead optimize equivalent dual expressions of the underlying 
information-theoretic objectives.

The dual-formulation perspective traces back to the convex duality of the RD problem~\cite{dembo2002source}. 
Specifically, by examining the optimality condition of the RD problem with respect to the conditional distribution $P_{Y|X}$, one can eliminate $P_{Y|X}$ and obtain an equivalent formulation that depends only on the marginal distribution $Q_Y$. 
The optimality condition is given by
\begin{equation}
    P_{Y|X=x}^\star(y)
    = \frac{Q_Y(y)\,e^{-\beta d(x,y)}}{\mathbb{E}_{Q_Y}[e^{-\beta d(x,Y)}]},
    \label{eq:rd-optimality}
\end{equation}
which directly characterizes the relation between the optimal conditional and the marginal distribution. 
Substituting this expression back into the RD objective yields a single-variable functional in $Q_Y$, from which a dual representation of the rate function can be derived.

\begin{theorem}
[Rate Function Duality~{\cite[Sec.~2]{dembo2002source}}]
The RD function admits the min–max form
\begin{equation}
    R(D)
    = \inf_{Q_Y}
    \sup_{\beta \ge 0}
    \Big\{
        -\beta D
        - \mathbb{E}_{P_X}
        \!\big[\log \mathbb{E}_{Q_Y}
        [e^{-\beta d(X,Y)}]\big]
    \Big\}.
    \label{eq:rate-minmax}
\end{equation}
\end{theorem}

For simplicity, we consider the fixed-$\beta$ case, where the objective reduces to minimizing the following functional:
\begin{equation}
    F(Q_Y)
    = - \mathbb{E}_{P_X}
        \!\big[\log \mathbb{E}_{Q_Y}
        [e^{-\beta d(X,Y)}]\big].
    \label{rd-dual}
\end{equation}

\textcolor{black}{This formulation offers a key advantage: once samples from $P_X$ are available and $Q_Y$ is parameterized, the functional $F(Q_Y)$ can be efficiently estimated via Monte Carlo approximation. 
The remaining challenge is how to model and sample from the unknown reconstruction distribution $Q_Y$~\cite{lei2022neural}. 
Motivated by recent advances in probabilistic generative modeling~\cite{goodfellow2014generative}, the \textit{Neural Estimation of the Rate–Distortion Function} (NERD)~\cite{lei2022neural} directly parameterizes the reconstruction marginal $Q_Y$ using a neural generator, thereby reducing the functional optimization over distributions to a tractable parameter optimization problem.}

Concretely, NERD models $Q_Y$ as the pushforward of a simple latent distribution $P_Z$ (e.g., a standard Gaussian) through a neural mapping $G_\theta : \mathcal{Z} \!\to\! \mathcal{Y}$, denoted as $Q_Y =  G_{\theta \#} P_Z$. By substituting this implicit representation into the dual functional defined in \eqref{rd-dual}, the problem is transformed into an optimization over $\theta$:
\begin{equation}
    \min_\theta\widehat{F}(\theta)
    = - \mathbb{E}_{P_X}
    \!\Big[\log \mathbb{E}_{P_Z}
    \big[e^{-\beta d(X,G_\theta(Z))}\big]\Big].
    \label{eq:nerd-population}
\end{equation}

Given a dataset $\mathcal{D} = \{x_i\}_{i=1}^n \!\sim\! P_X$, the empirical dual RD objective is defined by approximating the ensemble expectation over $X$ with the empirical average:
\begin{equation}
    \widehat{F}_n(\theta)
    = - \frac{1}{n}\sum_{i=1}^{n}
        \log \left( \mathbb{E}_{P_Z}
        \big[e^{-\beta d(x_i,G_\theta(Z))}\big] \right).
    \label{rd-dual-d}
\end{equation}

From a theoretical perspective, this neural estimation framework possesses strong consistency properties. Under mild regularity conditions regarding the alphabets and the distortion measure, the neural estimator is guaranteed to converge to the true RD function, as detailed in the following theorem.

\begin{theorem}[Strong Consistency of NERD~{\cite[Thm.~1]{lei2022neural}}]

\label{thm:nerd-consistency}
Suppose the source and reconstruction alphabets satisfy $\mathcal{X} = \mathcal{Y} = \mathbb{R}^m$, 
the latent prior $P_Z$ is supported on $\mathbb{R}^\ell$ and absolutely continuous with respect to the Lebesgue measure, 
and the distortion function $d(x,y)$ is $L_d$-Lipschitz continuous in both arguments.  
Then, the NERD
\[
    \hat{R}_\Theta(D)_n
    \!=\! \inf_{\theta \in \Theta}
    \sup_{\beta \ge 0}
    \Bigg\{\!
        \!-\beta D
        \!-\! \frac{1}{n}\!
        \sum_{i=1}^{n}
        \log
        \mathbb{E}_{P_Z}
        \!\big[
            e^{-\beta d(x_i, G_\theta(Z))}
        \big]\!
    \Bigg\}
\]
is \textit{strongly consistent}, i.e.,
\[
    \lim_{n \to \infty}
    \hat{R}_\Theta(D)_n
    = R(D),
    \quad \text{almost surely.}
\]
\end{theorem}

\textcolor{black}{Algorithmically, the training process alternates between data sampling and parameter updates. The complete training procedure is outlined as follows:
\begin{enumerate}[label=(\roman*)]
    \item Sample a mini-batch of source data $\{x_i\}_{i=1}^n$ independently from $P_X$, and a batch of latent variables $\{z_j\}_{j=1}^m$ from the prior distribution $P_Z$ (typically a standard Gaussian).
    \item Compute the empirical objective $\widehat{F}_n(\theta)$ according to~\eqref{rd-dual-d}.
    \item Update the parameters $\theta$ via SGD:
    \begin{equation*}
        \theta \leftarrow \theta - \eta\,\nabla_\theta \widehat{F}_{n}(\theta),
    \end{equation*}
    where $\eta$ denotes the learning rate.
\end{enumerate}}

Once the generator $G_\theta$ is trained, the RD pair $(R, D)$ can be efficiently estimated. For a fixed Lagrange multiplier $\beta$, the expected distortion is approximated empirically by:
\begin{equation*}
    D \approx \mathbb{E}_{P_X \otimes Q^\theta_Y}
    \left[
        d(X,Y)
        \frac{
            \exp(-\beta d(X,Y))
        }{
            \mathbb{E}_{Y' \sim Q^\theta_Y}
            [\exp(-\beta d(X,Y'))]
        }
    \right],
\end{equation*}
and the corresponding rate is derived using the relation $R = \widehat{F}(\theta) - \beta D$.

\begin{remark}
 {The NERD formulation here restates the original method in~\cite{lei2022neural} with a streamlined focus on the dual framework: the original algorithm estimates the RD function for fixed $D$ (using bisection search over $\beta$ per iteration to enforce the distortion constraint), while this presentation adopts a fixed-$\beta$ setup to directly optimize the dual functional, for a clearer analytical form and more stable training.}
\end{remark}

\textcolor{black}{Starting from the RD dual formulation, the IB objective can be obtained by replacing the distortion measure with the relevance-based distortion. 
Substituting this into~\eqref{rd-dual} yields the functional
\begin{equation*}
    \mathcal{F}(Q_T, R_{Y|T})
    = - \mathbb{E}_{P_X}
    \Big[
        \log \mathbb{E}_{Q_T}
        \big[
            e^{\beta \mathbb{E}_{P_{Y|X}}[\log R_{Y|T}(Y|T)]}
        \big]
    \Big],
\end{equation*}
which depends on an auxiliary marginal distribution $Q_T$ and a decoder $R_{Y|T}$.}

\textcolor{black}{Building on this formulation, the Mapping Approach Information Bottleneck (MAIB)~\cite{10806985} represents $Q_T$ as the pushforward of a latent prior $P_Z$ through the lens of the mapping approach~\cite{rose1994mapping}, thereby reducing the two-variable optimization over $(Q_T, R_{Y|T})$ to a single-variable problem over a conditional distribution $R_{Y|Z}$. 
The resulting objective takes the form
\begin{equation*}
    G(R_{Y|Z})
    = - \mathbb{E}_{P_X}
    \Big[
        \log \mathbb{E}_{P_Z}
        \big[
            e^{\beta \mathbb{E}_{P_{Y|X}}[\log R_{Y|Z}(Y|Z)]}
        \big]
    \Big],
\end{equation*}
which is optimized in practice by parameterizing $R_{Y|Z}$ with a neural network and applying SGD with Monte Carlo approximation.}

\textcolor{black}{Similarly, the iRD dual formulation is obtained by replacing the distortion function in the RD dual objective with the reduced distortion. 
In data-driven scenarios, however, the reduced distortion $\bar{d}(x,y)$ is typically unknown, as it depends on the intractable posterior $P_{S|X}$. 
The Neural Estimation of Indirect Rate--Distortion (NEIRD)~\cite{11161439} addresses this issue by introducing a neural estimator $g_{\hat\theta}(x,y)$ to approximate $\bar{d}(x,y)$. 
Leveraging the Markov structure $S \!\leftrightarrow\! X \!\leftrightarrow\! Y$, this approximation is learned via the minimum mean-square error (MMSE) objective:
\begin{equation}
    \mathcal{L}_{\text{MMSE}}(\hat\theta)
    = \mathbb{E}\big[(d(S,Y) - g_{\hat\theta}(X,Y))^2\big].
\end{equation}}

\textcolor{black}{With the learned distortion proxy $g_{\hat\theta}$, the dual objective becomes
\begin{equation}
    \widehat{F}_{iRD}(\theta,\hat\theta)
    = - \mathbb{E}_{P_X}
    \Big[
        \log \mathbb{E}_{P_Z}
        \big[
            e^{-\beta g_{\hat\theta}(X, T_\theta(Z))}
        \big]
    \Big],
\end{equation}
where the marginal $Q_Y$ is parameterized by a neural generator $T_\theta$. 
In practice, the model is trained by jointly optimizing both the generator and the distortion estimator.}

\textcolor{black}{\begin{remark}[Energy-based parameterization]
Within the dual-formulation framework, an alternative class of methods ~\cite{li2024rate,wu2025estimating} parameterizes the marginal distribution using energy-based models. 
Specifically, the marginal $Q_Y$ is modeled as a Boltzmann distribution $Q_Y^\theta(y) \propto e^{-E_\theta(y)}$. 
This provides a distinct advantage: while only the marginal $Q_Y$ is explicitly parameterized, the corresponding RD-optimal conditional $P_{Y|X}$ is implicitly induced through the same energy function. 
However, such models involve intractable normalization constants, making likelihood evaluation and direct optimization difficult. 
To tackle this issue, one instead derives a tractable gradient expression of the objective in a difference-of-expectations form, which can be estimated via sampling-based techniques such as Langevin MCMC~\cite{parisi1981correlation}.
\end{remark}}

\section{Experimental Comparisons and Discussions}

In this section, we conduct an experimental comparison of representative LtC methods for RD-type problems in several canonical settings where theoretical reference curves are available or commonly used. \textcolor{black}{Based on these empirical results and theoretical investigations, we compare the behaviors of different method families and highlight their respective strengths and limitations across various regimes.}

\subsection{Experimental Results}

Our experimental comparison begins with the RD problem on jointly Gaussian sources, following the setup of~\cite{lei2022neural}. The source distribution is chosen as a multivariate Gaussian $P_X=\mathcal{N}(0,\Sigma)$, where the covariance matrix admits the decomposition $\Sigma = U \mathrm{diag}(\sigma_1^2,\ldots,\sigma_d^2) U^T$ with $U$ drawn as a random orthogonal matrix. The eigenvalues are set to $\sigma_i = 2^{-i/10}$, ensuring that the RD curve has a known closed-form expression~\cite{berger2003rate}.

\begin{figure}[h]
    \centering
    \includegraphics[width=0.8\linewidth]{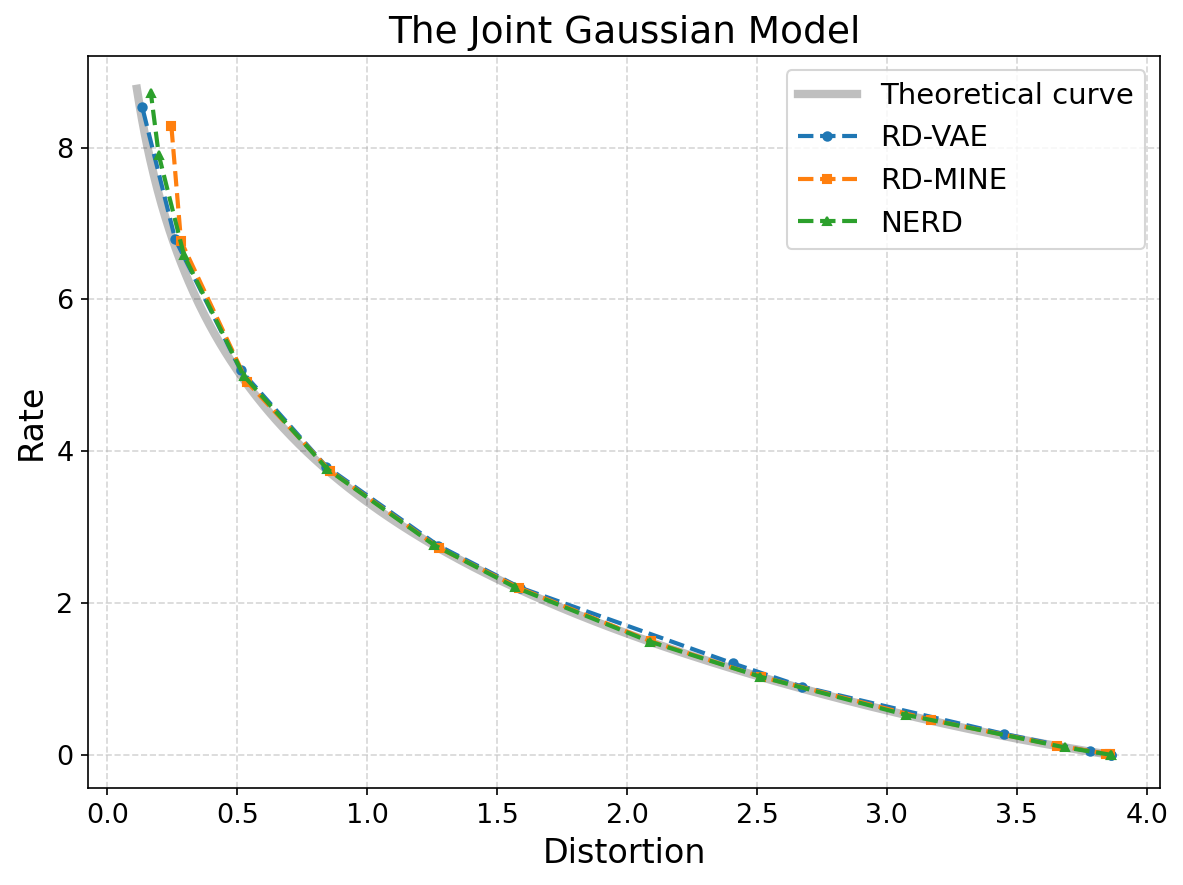}
    \caption{Comparison of neural RD estimators on a jointly Gaussian source. The theoretical RD curve is shown together with the estimates produced by RD-VAE, RD-MINE, and NERD.}
    \label{fig:gaussian_rd}
\end{figure}

Figure~\ref{fig:gaussian_rd} shows the RD curves for a 5-dimensional Gaussian source obtained using RD-VAE~\cite{yang2022towards}, RD-MINE~\cite{tsur2023rate}, and NERD~\cite{lei2022neural}. \textcolor{black}{For fairness, all methods are evaluated under comparable settings, including the same batch size (set to 12,800), similar network capacities, identical data distributions.} As shown in the figure, all methods are able to reproduce the shape of the theoretical RD function across the evaluated range. Minor discrepancies appear in the high-rate region, where RD-MINE and NERD exhibit increasing deviation from the reference curve, whereas RD-VAE remains comparatively stable across rates.

{RD-VAE's accuracy \textcolor{black}{is attributed to the property that}, for a Gaussian source under squared-error distortion, the optimal reconstruction distribution is also Gaussian. With Gaussian priors and posteriors, its variational family contains the optimal solution, yielding tight and accurate estimates over the entire rate range.}

In contrast, both MINE--based and dual-formulation approaches suffer from degraded high-rate performance due to a shared {difficulty} in estimating log-expectation terms. {Specifically, MINE-based methods rely on the Donsker--Varadhan objective, which involves estimating expressions of the form $\log \mathbb{E}[e^{T_\theta}]$, while dual-formulation methods require evaluating terms of the form $\log \mathbb{E}[e^{-\beta d(X,Y)}]$. In both cases, these log-expectation expressions are dominated by exponential weights, whose accurate estimation becomes increasingly challenging as the rate increases, often leading to biased or high-variance estimates under limited sampling~\cite{belghazi2018mutual,lei2022neural}. This challenge may potentially be alleviated by adopting more appropriate weighting or sampling strategies to better control estimator variance~\cite{owen2013monte }.}

The second experiment considers the IB problem on the MNIST dataset, a standard benchmark for image classification consisting of $28\times 28$ grayscale digit images and their corresponding digit labels. Here $X$ denotes the input image and $Y$ the class label. Since the label is taken as a deterministic function of the input image, the conditional distribution $P_{Y|X}$ collapses to a point mass, and consequently, the theoretical IB curve reduces to a simple piecewise-linear form determined by the entropy of the labels~\cite{kolchinsky2018caveats}.

\begin{figure}[h]
    \centering
    \includegraphics[width=0.8\linewidth]{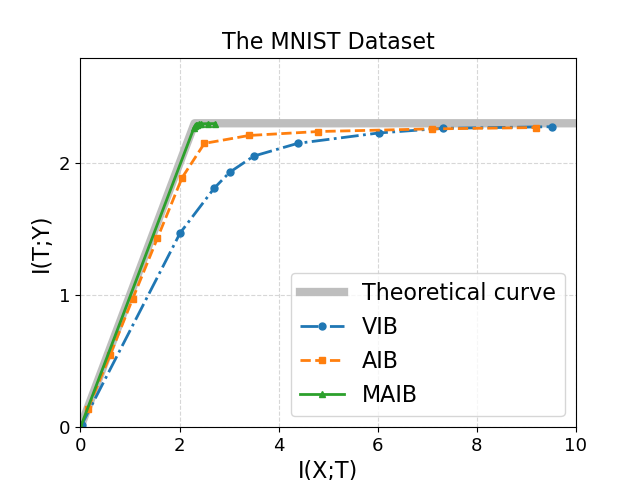}
    \caption{Comparison of neural IB estimators on the MNIST dataset. The theoretical IB curve is shown together with the estimates from VIB, AIB, and MAIB.}
    \label{fig:mnist_ib}
\end{figure}

Figure~\ref{fig:mnist_ib} displays the IB curves produced by VIB~\cite{alemi2016deep}, AIB~\cite{zhai2022adversarial}, and MAIB~\cite{10806985}, all trained with a batch size of 12,80. Among the three methods, MAIB aligns most closely with the theoretical piecewise-linear curve. Both AIB and VIB remain below the theoretical curve, with AIB achieving higher accuracy than VIB. Notably, the points produced by MAIB tend to cluster near the inflection point $(H(Y), H(Y))$, because the Lagrange multiplier $\beta$ is fixed during training~\cite{10806985}.

{The limitations of variational methods become evident beyond simple synthetic settings such as the Gaussian source. 
On real-world datasets such as MNIST, the true posterior and marginal distributions of the bottleneck representation are typically complex. The auxiliary distributions (e.g., $ Q_T $ , $ R_{Y|T}$) employed in variational approximations often fail to capture this complexity,  resulting in looser bounds and biased estimates of information quantities~\cite{poole2019variational}.}

The third experiment focuses on the iRD problem. We present the NEIRD results on the high-dimensional Gaussian model from~\cite{11161439}. Here, the observation $X$ follows $\mathcal{N}(0,K_X)$ with $K_X = 2I_{120}$, while the source $S$ is generated via the linear model $S = HX + W$, where $H$ is a sparse $6 \times 120$ matrix whose entries take values in $\{-1,0,1\}$ with probabilities $0.05$, $0.90$, and $0.05$, and the noise satisfies $W \sim \mathcal{N}(0, I_6)$.

\begin{figure}[h]
    \centering
    \includegraphics[width=0.49\linewidth]{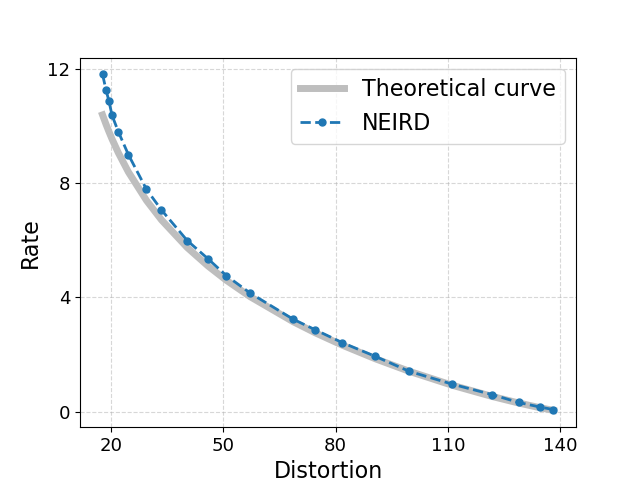}
    \includegraphics[width=0.49\linewidth]{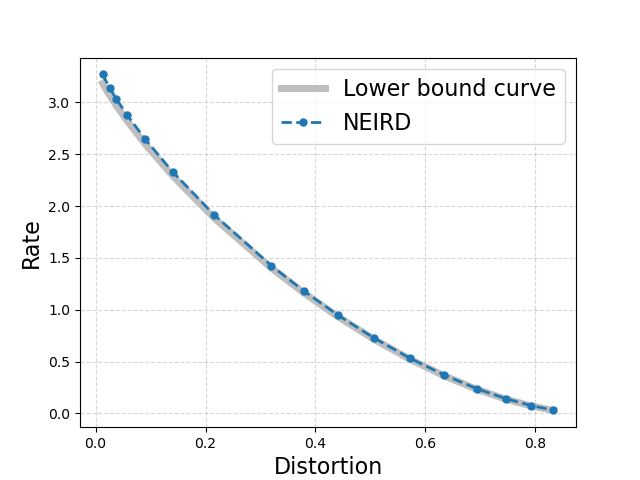}
    \caption{NEIRD results on (left) the joint Gaussian model and (right) the MNIST dataset under Hamming distortion. \textcolor{black}{In the Gaussian case, the learned iRD curve is compared with the theoretical iRD function. For MNIST, it is compared with the corresponding analytical lower bound.}}
    \label{fig:neird} 
\end{figure}

We also include the NEIRD results on MNIST under the Hamming distortion, where the distortion measure $\bar{d}$ admits a closed-form expression and the theoretical iRD function admits an analytical lower bound. Figure~\ref{fig:neird} shows the resulting iRD curves for the Gaussian and MNIST settings, reproduced from~\cite{11161439}. \textcolor{black}{For the Gaussian model, NEIRD closely matches the theoretical iRD curve across most of the distortion range; for MNIST, the learned curve remains close to the analytical lower bound.} In both cases, small but observable deviations appear in the high-rate region, consistent with the limitations of dual-formulation methods discussed earlier.

\subsection{Comparative Analysis of LtC Methods}

\begin{table*}[t]
\centering
\caption{Comparison of LtC Methods for RD-Type Problems}
\label{tab:ltc_comparison}
\normalsize
\renewcommand{\arraystretch}{1.35}

\begin{tabularx}{\textwidth}{
>{\raggedright\arraybackslash}p{2.5cm}
>{\raggedright\arraybackslash}X
>{\raggedright\arraybackslash}X
>{\raggedright\arraybackslash}X
}
\toprule
\textbf{Aspect} 
& \textbf{Variational Methods} 
& \textbf{MINE-Based Methods} 
& \textbf{Dual-Form Methods} \\
\midrule

\textbf{Optimization structure} 
& Single-stage minimization
& Min--max optimization with multiple critic updates 
& Single objective involving pairwise distortion computations \\

\textbf{Per-iteration complexity} 
& $\mathcal{O}(B)$ 
& $\mathcal{O}(KB)$
& $\mathcal{O}(BM)$ \\

\textbf{Primary bias source} 
& Model bias from restricted variational family 
& Estimation bias from log-expectation terms 
& Estimation bias from log-expectation terms \\

\textbf{Training stability} 
& High; stable single-objective optimization 
& Low; adversarial training with high-variance gradients 
& Medium; no adversarial training but numerically sensitive \\

\textbf{Suitable regimes} 
& Large-scale settings requiring efficient and scalable optimization 
& Settings where mutual information estimation is essential and sufficient data is available 
& Low-rate regimes where estimation variance is better controlled  \\

\bottomrule
\end{tabularx}
\end{table*}

\textcolor{black}{We compare variational approaches, neural mutual-information estimation methods, and dual-form approaches from three key perspectives: computational complexity, sources of bias, and practical stability. The main findings of this comparison are summarized in Table~\ref{tab:ltc_comparison}.}

\textcolor{black}{\textbf{Computational complexity.}
Let $B$ denote the mini-batch size. Variational methods are computationally the most efficient, as each iteration involves a single-stage minimization with standard forward and backward passes, leading to a per-iteration cost that scales linearly with $B$.
In contrast, MINE-based methods introduce an additional critic network and require a min--max optimization procedure, resulting in an effective complexity of $\mathcal{O}(K B)$ due to multiple critic updates per iteration, where $K$ denotes the number of critic updates.
Dual-form methods can become computationally expensive in practice, as each iteration requires evaluating pairwise distortions between $B$ data samples and $M$ samples drawn from the learned distribution, resulting in a complexity of $\mathcal{O}(BM)$; when $M$ scales proportionally with $B$ (i.e., $M \sim B$), this becomes quadratic $\mathcal{O}(B^2)$.}

\textcolor{black}{\textbf{Sources of bias.}
The primary source of bias in variational methods is the variational gap. Since the true optimal distributions are approximated within a restricted variational family, the resulting objective typically constitutes an upper bound, which can be loose when the model class lacks sufficient expressiveness.
In contrast, both MINE-based and dual-form methods suffer from degraded performance in the high-rate regime due to a shared difficulty of estimating log-expectation terms. These expressions are dominated by exponentially weighted samples, making accurate estimation challenging under finite samples and leading to high-variance estimates.}

\textcolor{black}{\textbf{Stability and practical regimes.}
Variational methods are generally the most stable, benefiting from single-objective optimization and well-established training procedures. They are particularly suitable for large-scale applications that require efficient and scalable optimization. 
MINE-based methods are typically less stable due to their adversarial structure and high-variance gradient estimates. They are advantageous when direct optimization of mutual information is essential and sufficient data and computational resources are available. 
Dual-form methods offer a compromise: they avoid min--max optimization but may still suffer from numerical instability associated with log-expectation estimation. They are often more effective in low-rate regimes, where estimation variance is better controlled.}

\section{Conclusion and Outlook}

This survey has reviewed recent progress in LtC methods for evaluating the RD-type problems. By categorizing existing methods into three families—variational approaches, neural mutual-information estimators, and dual-form methods—we clarified their computational mechanisms, theoretical underpinnings, and practical trade-offs. Empirical studies on canonical benchmarks have validated the correctness and practical feasibility of existing LtC methods, showing that they can reliably reproduce theoretical reference curves or capture the expected behaviors in representative settings. \textcolor{black}{However, these experiments also reveal inherent limitations: variational methods suffer from variational gap when the true distribution is complex, whereas MINE-based and dual-form methods tend to degrade in the high-rate regime due to the intrinsic difficulty of estimating log-expectation terms.} {Collectively, these limitations indicate that, despite their demonstrated effectiveness on standard benchmarks,  challenges remain for current LtC methods when applied to large-scale datasets, which point to promising directions for future research and development.}

Building on these insights, we outline three directions for future research. First, leveraging more expressive neural network models—such as diffusion models~\cite{ho2020denoising} and score-based generative frameworks~\cite{song2021score}—can enhance the ability of neural estimators to capture complex data geometries, potentially tightening the approximation of the RD curves. Second, bridging information-theoretic computation with statistical physics offers a novel lens: drawing parallels between RD problem and free-energy minimization~\cite{rose1994mapping}, and leveraging tools like Gibbs measures and phase transition analysis~\cite{georgii2011gibbs}, may yield more stable numerical schemes and deeper insights into optimal representation learning.  {Third, developing more effective statistical weighting and sampling strategies  to better control estimator variance can help reduce estimation bias and variance~\cite{owen2013monte}, thereby improving the reliability and scalability of LtC methods in high-dimensional regimes.}

In summary, the LtC paradigm provides a  {viable} route to bringing RD objectives into learning pipelines by casting them as differentiable, data-driven optimization problems. 
Continued progress in expressive probabilistic modeling, numerical optimization, and information theory will be essential for improving accuracy and robustness in high-dimensional regimes, and for expanding the impact of RD framework in learned compression, representation learning, and high-dimensional signal processing.

\bibliographystyle{IEEEtran}
\bibliography{ref}

\end{document}